# The unitary dependence theory for understanding quantum circuits and states

Zixuan Hu and Sabre Kais*

*Department of Chemistry, Department of Physics, and Purdue Quantum Science and Engineering Institute, Purdue University, West Lafayette, IN 47907, United States*
*Email:* kais@purdue.edu

Abstract: We develop a unitary dependence theory to characterize the behaviors of quantum circuits and states in terms of how quantum gates manipulate qubits and determine their measurement probabilities. A qubit has dependence on a 1-qubit unitary gate if its measurement probabilities depend on the parameters of the gate. A 1-qubit unitary creates such a dependence onto the target qubit, and a CNOT gate copies all the dependences from the control qubit to the target qubit. The complete dependence picture of the output state details the connections qubits may have when being manipulated or measured. Compared to the conventional entanglement description of quantum circuits and states, the dependence picture offers more practical information, easier generalization to many-qubit systems, and better robustness upon partitioning of the system. The unitary dependence theory is a useful tool for understanding quantum circuits and states that is based on the practical ideas of how manipulations and measurements are affected by elementary gates.

1. Introduction

Over several decades, the search for quantum algorithms with efficient quantum circuit implementations has resulted in numerous theoretical and technological advancements [1-13]. Notable quantum algorithms with the potential to outperform classical algorithms include: the phase estimation algorithm [14], Shor's factorization algorithm [15], the Harrow-Hassidim-Lloyd algorithm for linear systems [16], the hybrid classical-quantum algorithms [17, 18], the quantum machine learning algorithms [19-21], and quantum algorithms for open quantum dynamics [22-24].

Despite the expansive selection of ideas, most existing quantum algorithms are either discovered accidentally or adapted from classical algorithms, and a systematic way to understand and design novel quantum algorithms with efficient quantum circuit implementation is currently lacking. Considering the enormous search space of quantum circuits, we need a general theory to classify all quantum circuits and states, and then study their behaviors based on the classification. To this end, we have proposed a theory to classify all quantum circuits of a given length into finite types based on the "qubit functional configuration" [25]. In this work we proceed to develop a *unitary dependence theory* to characterize the behaviors of quantum circuits and states in terms of how quantum gates manipulate qubits and determine their measurement probabilities. Any quantum circuit can be decomposed into a sequence of elementary gates that include 1-qubit unitaries and

2-qubit CNOT gates. The initial state is transformed by the circuit into the final state which then undergoes projection measurements in the computational basis $\{|0\rangle,|1\rangle\}$. The probability of measuring $|0\rangle$ versus $|1\rangle$ on each qubit then determines the measurement statistics of the final state which can be used to characterize the behavior of the quantum circuit and the final state. In this work we study how the probabilities of measuring each qubit are affected by elementary gates. The basic rules are: 1. a 1-qubit unitary $U_k$ makes the target qubit $q_k$'s measurement probabilities depend on $U_k$; 2. a CNOT gate $CX_{j \to k}$ copies all the control qubit $q_j$'s dependences to the target qubit $q_k$. By these rules, each qubit $q_k$ at the final state has its own collection of dependences that may be created on itself by some $U_k$s or copied to it from other qubits by some CNOT gates. While a dependence may be copied from other qubits, it must have been created by some $U_i$ on another qubit $q_i$ – i.e. any given dependence can be traced back to the original 1-qubit unitary that has created it in the first place – therefore a complete *dependence picture* of which qubit depends on which 1-qubit unitaries can be drawn by analyzing the gate sequence of a quantum circuit. The *dependence picture* carries important information that helps us understand the behaviors of quantum circuits. For example, if two qubits share dependences on some 1-qubit unitaries, then their measurement probabilities must be dependent; while if they do not share any dependence, their measurement probabilities must be independent. From the circuit design perspective, varying the parameters of a 1-qubit unitary shared by multiple qubits allows us to manipulate the measurement probabilities of all involved qubits together; while qubit-specific manipulations have to be implemented with 1-qubit unitaries that are unique to the particular qubit. The dependence picture is a novel perspective for understanding quantum circuits and states because it is distinct from the conventional way of using entanglement as a descriptor of complexity [26]. Compared to the abstract formalism of multi-qubit entanglement [27-29], the dependence picture is more directly connected to the practicalities of using parameterized quantum gates to manipulate qubits and create desirable measurement statistics in the output states. Furthermore, in a deeper discussion of the unitary dependence theory, we find that under certain conditions the dependences originated from the same unitary source can cancel when duplicated on the same qubit, which reduces complexity and simplifies the dependence picture. Interestingly, in studying the cancellability of dependences, we find that entanglement can protect the cancellability from getting broken by local 1-qubit unitaries, and thus an intricate relation between the unitary dependence and entanglement exists. Finally the theory has been applied to the widely-used hardware-efficient ansatz [30-33] to demonstrate its ability to characterize the behaviors of different ansatzes.

## 2. Theory of the unitary dependence of qubits

**2.1 Basic rules for the creation and copy of unitary dependences.** The core idea of quantum computing is to use a small number of qubits to perform computation on the state space of a huge dimension that scales exponentially with the qubit number *n*. Qubits are therefore the fundamental targets for quantum gates and measurements. Thus to develop a theory to systematically understand quantum circuits we need to focus on how quantum gates affect qubits. Any arbitrary quantum circuit takes in an initial input state, applies a sequence of elementary gates to transform it into a final state, and then measures all or selective qubits in the computational basis. Without loss of generality, we start with the simplest 3-qubit initial state $|000\rangle_{123}$ (where the subscripts 123 denote the qubit number identifiers), then all three qubits start with probability 1 for measuring $|0\rangle$. Now if we apply an arbitrary 1-qubit unitary $U_1(a_1,a_2,\alpha) = \begin{pmatrix} a_1 & a_2^* e^{i\alpha} \\ a_2 & -a_1^* e^{i\alpha} \end{pmatrix}$ (with $|a_1|^2 + |a_2|^2 = 1$) to $q_1$ then the state becomes $(a_1|0\rangle_1 + a_2|1\rangle_1) \otimes |00\rangle_{23}$ and measuring $q_1$ will yield the probabilities $p(|0\rangle_1) = |a_1|^2$ and $p(|1\rangle_1) = |a_2|^2$:

> **Definition 1**: The situation that the probabilities of measuring a qubit $q_k$ depend on the parameters of a 1-qubit unitary $U_k$ is defined as "$q_k$ has acquired the dependence on $U_k$" or alternatively "$U_k$'s dependence has been created on $q_k$".

By the definition $q_1$ has acquired the dependence on $U_1(a_1,a_2,\alpha)$ in the above example. Note in this particular example the phase $\alpha$ has no obvious effect on $p(|0\rangle_1)$ or $p(|1\rangle_1)$, but in general it may affect the probabilities in certain cases. Now if we apply another 1-qubit unitary $U_2(b_1,b_2,\beta) = \begin{pmatrix} b_1 & b_2^* e^{i\beta} \\ b_2 & -b_1^* e^{i\beta} \end{pmatrix}$ (with $|b_1|^2 + |b_2|^2 = 1$) to $q_2$ then the state becomes $(a_1|0\rangle_1 + a_2|1\rangle_1) \otimes (b_1|0\rangle_2 + b_2|1\rangle_2) \otimes |0\rangle_3$, and $q_2$ has acquired the dependence on $U_2(b_1,b_2,\beta)$. Now if we apply the CNOT gate $CX_{1\to 2}$ with the subscript $1\to 2$ meaning $q_1$ is the control and $q_2$ is the target, then the state becomes $[a_1|0\rangle_1 \otimes (b_1|0\rangle_2 + b_2|1\rangle_2) + a_2|1\rangle_1 \otimes (b_2|0\rangle_2 + b_1|1\rangle_2)] \otimes |0\rangle_3$, and now measuring $q_2$ will yield $p(|0\rangle_2) = |a_1 b_1|^2 + |a_2 b_2|^2$: we see that the probabilities of measuring $q_2$ now depend on $U_1(a_1,a_2,\alpha)$ too and thus $q_2$ has acquired the dependence on $U_1(a_1,a_2,\alpha)$ through $CX_{1\to 2}$. It is obvious that the control qubit $q_1$ is unaffected by $CX_{1\to 2}$. Next if we apply another CNOT gate $CX_{2\to 3}$, then the state becomes $(a_1 b_1|0\rangle_1 + a_2 b_2|1\rangle_1)|00\rangle_{23} + (a_1 b_2|0\rangle_1 + a_2 b_1|1\rangle_1)|11\rangle_{23}$, and measuring $q_3$ will yield $p(|0\rangle_3) = |a_1 b_1|^2 + |a_2 b_2|^2$ such that $q_3$ has acquired all the dependences on $q_2$ – both $U_1(a_1,a_2,\alpha)$ and $U_2(b_1,b_2,\beta)$ – through $CX_{2\to 3}$. To summarize the entire process, we started with $|000\rangle_{123}$ having no dependence on any 1-qubit unitary for any qubit, created $U_1(a_1,a_2,\alpha)$'s dependence on $q_1$, created $U_2(b_1,b_2,\beta)$'s dependence on $q_2$, copied $q_1$'s dependence to $q_2$ by $CX_{1\to 2}$, and copied $q_2$'s dependences to $q_3$ by $CX_{2\to 3}$.

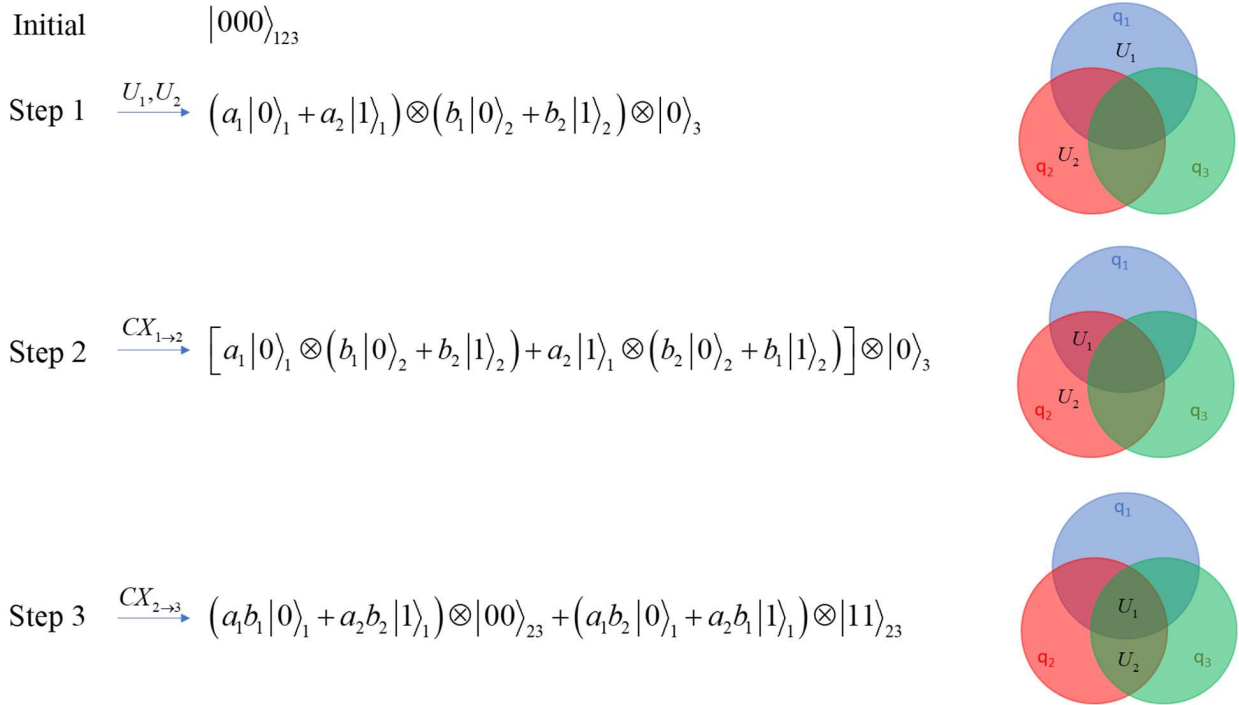

Initial $\quad |000\rangle_{123}$

Step 1 $\xrightarrow{U_1, U_2} (a_1|0\rangle_1 + a_2|1\rangle_1) \otimes (b_1|0\rangle_2 + b_2|1\rangle_2) \otimes |0\rangle_3$

Step 2 $\xrightarrow{CX_{1\to 2}} [a_1|0\rangle_1 \otimes (b_1|0\rangle_2 + b_2|1\rangle_2) + a_2|1\rangle_1 \otimes (b_2|0\rangle_2 + b_1|1\rangle_2)] \otimes |0\rangle_3$

Step 3 $\xrightarrow{CX_{2\to 3}} (a_1 b_1|0\rangle_1 + a_2 b_2|1\rangle_1) \otimes |00\rangle_{23} + (a_1 b_2|0\rangle_1 + a_2 b_1|1\rangle_1) \otimes |11\rangle_{23}$

Figure 1. Showing the dependence creation and copy process with the state after each step described by a Venn diagram that shows how unitary dependences are shared among the qubits.

Figure 1 shows the entire process of dependence creation by 1-qubit unitaries and copy by CNOT gates. At each step, the dependence picture is illustrated by a Venn diagram that clearly shows how the unitary dependences are shared by the qubits. At the final state we have $U_2(b_1, b_2, \beta)$ shared by $q_2$ and $q_3$, and $U_1(a_1, a_2, \alpha)$ shared by all three qubits. The dependence picture clearly tells us that the probabilities of all three qubits are dependent such that they cannot be considered independent random variables when measured. Furthermore, changing the parameters $a_1$ and $a_2$ of $U_1(a_1, a_2, \alpha)$ will modify the probabilities of all three qubits, while changing $b_1$ and $b_2$ of $U_2(b_1, b_2, \beta)$ will modify only $q_2$ and $q_3$. As there is no 1-qubit unitary unique to any qubit, in this circuit we cannot specifically modify a single qubit.

Having seen a simple 3-qubit example, next we propose the rules for a general *n*-qubit system that:

> **Rule 1**: A 1-qubit unitary $U_k$ applied to the qubit $q_k$ creates its dependence on $q_k$ only;

> **Rule 2**: A CNOT gate $CX_{j\to k}$ copies all the control qubit $q_j$'s dependences to the target qubit $q_k$.

A special case of Rule 1 has been proven in Theorem 1 of our previous quantum encryption study [34] for real parameters in unitaries and states. Generalizing the proof to complex parameters is straightforward:

**Proof for Rule 1**: An arbitrary $n$-qubit state can be written as the Schmidt decomposition form with respect to any given $q_k$:

$$\left|\phi^{(n)}\right\rangle = C_1 \left|\phi_1^{(n-1)}\right\rangle \left(a_1 |0\rangle_k + a_2 |1\rangle_k\right) + C_2 \left|\phi_2^{(n-1)}\right\rangle \left(a_2^* |0\rangle_k - a_1^* |1\rangle_k\right) e^{i\alpha} \tag{1}$$

where $\left|\phi_1^{(n-1)}\right\rangle$ and $\left|\phi_2^{(n-1)}\right\rangle$ are orthogonal $(n-1)$-qubit states that exclude $q_k$, $C_1$ and $C_2$ are non-negative real numbers satisfying $|C_1|^2 + |C_2|^2 = 1$, $a_1$ and $a_2$ are complex numbers satisfying $|a_1|^2 + |a_2|^2 = 1$. Clearly the probability of measuring $|0\rangle$ on $q_k$ is $p(|0\rangle_k) = |C_1 a_1|^2 + |C_2 a_2|^2$ and thus we may generalize the definition of unitary dependence to say that $q_k$ has the dependences on the two pairs of coefficients $(C_1, C_2)$ and $(a_1, a_2)$. Both $(C_1, C_2)$ and $(a_1, a_2)$ can be very complicated as they may include the dependences of many 1-qubit unitaries that were used to create $|\phi^n\rangle$. However, because $|\psi_1\rangle = a_1 |0\rangle_k + a_2 |1\rangle_k$ and $|\psi_2\rangle = (a_2^* |0\rangle_k - a_1^* |1\rangle_k) e^{i\alpha}$ are orthogonal, $(a_1, a_2)$ is a *local dependence* that only applies to $q_k$ while $(C_1, C_2)$ is a *shared dependence* that $q_k$ may share with other qubits. Now if we apply $U_k = \begin{pmatrix} u_1 & u_2^* e^{i\theta} \\ u_2 & -u_1^* e^{i\theta} \end{pmatrix}$ on $q_k$, as $|U_k \psi_1\rangle$ is always orthogonal to $|U_k \psi_2\rangle$, the dependence on $U_k$ will be added to the local dependence and thus only $q_k$ will acquire it. This proves Rule 1 that a 1-qubit unitary $U_k$ applied to the qubit $q_k$ creates its dependence on $q_k$ only.

**Proof for Rule 2**: The action of a CNOT gate $CX_{j \to k}$ is to keep $q_k$ intact when $q_j = |0\rangle$ and bit-flip $q_k$ if $q_j = |1\rangle$ – this effectively calculates the binary sum $q_k \oplus q_j$ and stores its value on $q_k$ [25, 35]. When we measure $q_k$ after $CX_{j \to k}$, the probabilities of getting $|0\rangle$ and $|1\rangle$ are actually the probabilities of getting $q_k \oplus q_j = |0\rangle$ and $q_k \oplus q_j = |1\rangle$, thus all the 1-qubit unitaries that affect $q_j$'s measurement probabilities will now also affect $q_k$'s probabilities after $CX_{j \to k}$. In the meanwhile, it is obvious that $q_j$'s measurement probabilities are not affected by $CX_{j \to k}$. Therefore Rule 2 has been proven that "A CNOT gate $CX_{j \to k}$ copies all the control qubit $q_j$'s dependences to the target qubit $q_k$".

The unitary dependence theory allows us to generate dependence pictures as illustrated by the Venn diagrams in Figure 1. The dependence picture of a quantum circuit or state provides two important pieces of information that characterize its behaviors. The first piece is how the measurement probabilities of qubits depend on each other in the output state. If multiple qubits share dependences on a certain collection of 1-qubit unitaries, their measurement results will behave as dependent variables; if multiple qubits do not share any unitary dependences, their measurement results will behave as independent variables. The dependence picture thus allows us to characterize the measurement statistics of the output state, which is an important property of a quantum circuit or state [8]. The second piece is how the qubits can be manipulated together by varying parameters of the shared 1-qubit unitaries. If a 1-qubit unitary is shared by multiple qubits, varying its parameters can change the measurement probabilities of all involved qubits together, thus allowing collective manipulations of subgroups of qubits. On the other hand, if the manipulation of an individual qubit is needed, then the qubit must have 1-qubit unitaries not shared by any other qubits. This is a design principle that can guide the development of e.g. parameterized quantum circuits [17, 26, 36, 37] where circuits and ansatz states are tested by varying gate parameters.

**2.2 Advantages over the entanglement description of quantum circuits and states.** The unitary dependence theory and the dependence picture are distinct from the conventional way of using entanglement to understand quantum circuits and states. In this section we detail the differences between the two theories and show the advantages of the dependence picture over the entanglement understanding.

**2.2.1** Entanglement is an important subject in quantum physics that demonstrates quantum-only features that have no classical equivalent. However, from the perspective of quantum circuit design, entanglement is too abstract and not easily connected to the practicalities discussed above on how qubits are related in manipulations and measurement probabilities. To see this, consider the state:

$$|\psi\rangle = a_1\left(|00\rangle_{12} + a_2|11\rangle_{12}\right) \otimes b_1|0\rangle_3 + a_1\left(|01\rangle_{12} + a_2|10\rangle_{12}\right) \otimes b_2|1\rangle_3 \tag{2}$$

where $a_1$, $a_2$, $b_1$, $b_2$ are complex numbers satisfying $|a_1|^2 + |a_2|^2 = 1$ and $|b_1|^2 + |b_2|^2 = 1$. In $|\psi\rangle$, $q_1$ and $q_3$ are considered entangled because there is no separable way to write the two qubits as a product state. However, by simple inspection we have $p(|0\rangle_1) = |a_1|^2$ and $p(|0\rangle_3) = |b_1|^2$ – i.e. $q_1$ has dependence on the pair $(a_1, a_2)$ while $q_3$ has dependence on the pair $(b_1, b_2)$ – which means $q_1$ and $q_3$ have independent measurement probabilities and can be manipulated separately. So the entanglement picture fails to characterize the relation between $q_1$ and $q_3$. In the meanwhile, by the unitary dependence theory, $|\psi\rangle$ is created by the process shown in Figure 2 and the relation between $q_1$ and $q_3$ can be easily seen in the Venn diagram.

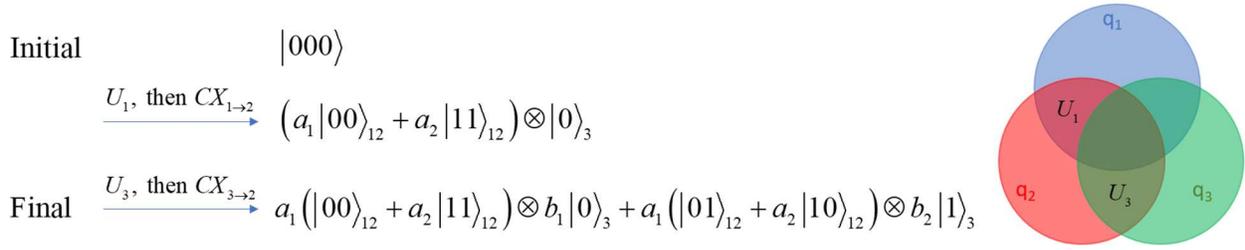

| Initial | | $\lvert 000\rangle$ |
|---|---|---|
| | $\xrightarrow{U_1,\ \text{then}\ CX_{1\to 2}}$ | $\left(a_1 \lvert 00\rangle_{12} + a_2 \lvert 11\rangle_{12}\right) \otimes \lvert 0\rangle_3$ |
| Final | $\xrightarrow{U_3,\ \text{then}\ CX_{3\to 2}}$ | $a_1\left(\lvert 00\rangle_{12} + a_2 \lvert 11\rangle_{12}\right)\otimes b_1\lvert 0\rangle_3 + a_1\left(\lvert 01\rangle_{12} + a_2 \lvert 10\rangle_{12}\right)\otimes b_2\lvert 1\rangle_3$ |

Figure 2. Showing the creation process of $\lvert\psi\rangle$ and the final Venn diagram. $U_1(a_1, a_2, \alpha)$ and $U_3(b_1, b_2, \beta)$ are 1-qubit unitaries applied to $q_1$ and $q_3$ respectively. Clearly $q_1$ and $q_3$ are independent in the dependence picture.

In Figure 2 we have $U_1(a_1, a_2, \alpha)$ shared by $q_1$ and $q_2$, and $U_3(b_1, b_2, \beta)$ shared by $q_2$ and $q_3$, therefore $q_1$ and $q_3$ are independent for both measurement and manipulation. This example clearly shows the dependence picture is fundamentally different from the entanglement picture because two entangled qubits may have no shared dependence and thus are considered to be independent by the dependence theory.

**2.2.2** Using entanglement to describe quantum circuits and states can become more confusing when considering multiple-qubit systems where many different measures and theories for multipartite entanglement exist [27-29]. For example, for a 3-qubit system there are the Greenberger–Horne–Zeilinger (GHZ) state $\lvert GHZ\rangle = \frac{1}{\sqrt{2}}(\lvert 000\rangle + \lvert 111\rangle)$ and the W state $\lvert W\rangle = \frac{1}{\sqrt{3}}(\lvert 001\rangle + \lvert 010\rangle + \lvert 100\rangle)$, which are well-known to be different and can be explained by various multipartite entanglement measures [27-29]. However, most multipartite entanglement measures generate some numbers for fixed values of parameters and thus cannot characterize parameterized quantum circuits and states: e.g. $a_1\lvert 000\rangle + a_2\lvert 111\rangle$ will produce different values of the von Neumann entropy, given different values for $a_1$ and $a_2$. In the meanwhile the dependence picture is naturally suited to the ideas of using finite structures with varying parameters to characterize large collections of quantum circuits and states together [25, 38]: e.g. $a_1\lvert 000\rangle + a_2\lvert 111\rangle$ can be described by one simple Venn diagram in Figure 3.

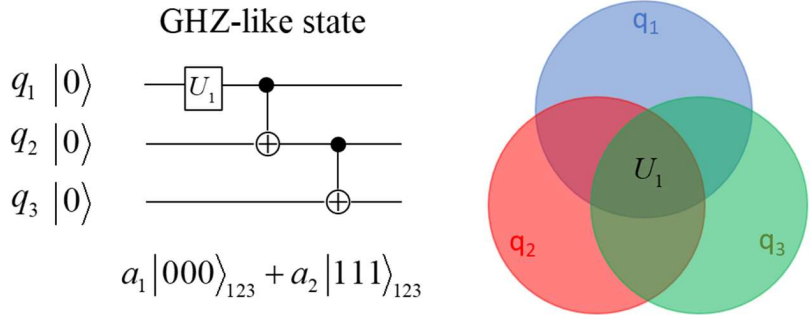

Figure 3. Showing the creation circuit of the GHZ-like state and the Venn diagram of the dependence picture. $U_1 = U_1(a_1, a_2, \alpha)$ is a 1-qubit unitary applied to $q_1$. When $a_1 = a_2 = \frac{1}{\sqrt{2}}$ the state becomes the GHZ state.

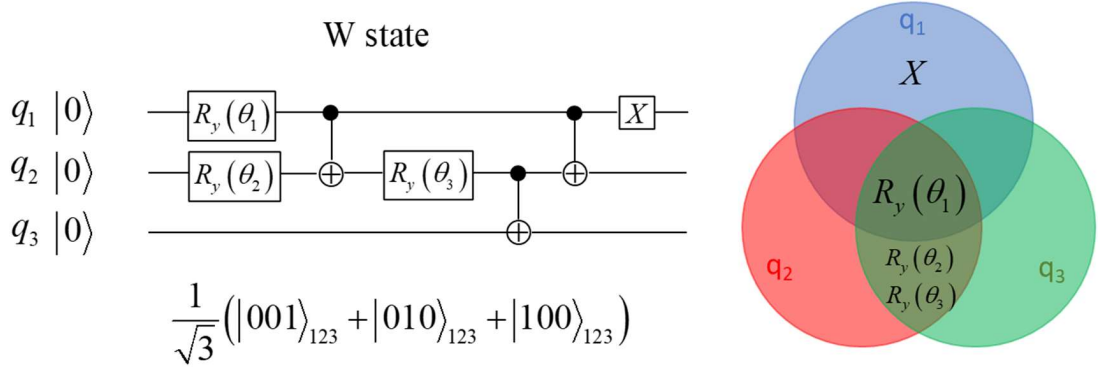

Figure 4. Showing the creation circuit of the W state and the Venn diagram of the dependence picture.

$$R_y(\theta) = \begin{pmatrix} \cos\frac{\theta}{2} & -\sin\frac{\theta}{2} \\ \sin\frac{\theta}{2} & \cos\frac{\theta}{2} \end{pmatrix}, \quad \theta_1 = 2\arccos\left(\frac{1}{\sqrt{3}}\right), \quad \theta_2 = -\theta_3 = \frac{\pi}{4}.$$

Although there is a theory to generalize the GHZ state and W state into the GHZ class and W class [28], the process of determining the class for an arbitrary state involves calculating the determinants of all the possible bipartite reduced density matrices and then the 3-tangle [27, 28]: this is a very complex process and more importantly not describing the qubit relation in measurement and manipulation. For example, it is not obvious whether the state $|\psi\rangle$ in Equation (2) belongs to the GHZ or W class, and even if we can determine its class, it has a dependence picture (see Figure 2) distinct from both the GHZ and W classes in Figure 3 and Figure 4 such that the entanglement class does not describe the qubit relation of interest in the current study. Furthermore, when the number of qubits increases beyond 3, the number of possible entanglement classes becomes infinite and there is no practical way to determine the class for an arbitrary state [28]. On the other hand, no matter how many qubits there are, the dependence picture can always be easily generated by going through the gate sequence associated with any quantum circuit or state, one gate at a time. So as long as the gate count is polynomial the process of determining the dependence picture is polynomial. When the number of qubits increases it may be difficult to draw the Venn diagram, but the dependence picture can still be easily described by listing the qubits with the unitary dependences they have, and listing the 1-qubit unitaries with the qubits they belong to. For example, the W state dependence picture in Figure 4 can be alternatively described by

The W state unitary dependence description by qubits $\left(q_1:\{R_y(\theta_1),X\},\ q_2:\{R_y(\theta_1),R_y(\theta_2),R_y(\theta_3)\},\ q_3:\{R_y(\theta_1),R_y(\theta_2),R_y(\theta_3)\}\right)$ (3) by 1-qubit unitaries $\left(R_y(\theta_1):\{q_1,q_2,q_3\},\ R_y(\theta_2):\{q_2,q_3\},\ X:\{q_1\},\ R_y(\theta_3):\{q_2,q_3\}\right)$ and this description always works for systems with more qubits.

**2.2.3** When studying the behavior of quantum circuits and states involving many qubits, there are cases where we want to temporarily focus on a subset of qubits and treat the other qubits as an averaging background. For entanglement, the qubits that are considered entangled in the total system may become unentangled in the subsystem. For example, the GHZ-like entangled state $a_1|000\rangle+a_2|111\rangle$ will become the reduced density matrix $\rho=a_1|00\rangle\langle00|+a_2|11\rangle\langle11|$ when any one of the qubits is removed, and this is clearly an unentangled (separable) mixed state [39]. This shows that entanglement is sensitive to how we partition the space into subsets of qubits. On the other hand, the dependence picture is robust regardless how we partition the space as we can take the Venn diagrams or the descriptions like in Equation (3), remove all the unwanted qubits, and the remaining parts will describe the correct relations among the qubits of interest. Again using the example of $a_1|000\rangle+a_2|111\rangle$ shown in Figure 3, removing any one of the qubits will yield the simple description of two qubits sharing one 1-qubit unitary: $\left(q_1:\{U_1\},\ q_2:\{U_1\}\right)$ and $\left(U_1:\{q_1,q_2\}\right)$, therefore the relation between the two qubits stays the same in the absence of $q_3$. This further illustrates the fact that the dependences of qubits on 1-qubit unitaries are more robust physical properties that are invariant upon partitioning of the system, while entanglement is less robust that may change upon partitioning of the system.

**2.3 The dependence cancelling rule and the entanglement protection of cancellability.**

**2.3.1** Having seen the basic rules of dependence creation and copy in Section 2.1, next we discuss the situation where dependences originated from the same unitary source can cancel when duplicated on the same qubit. In the simplest case, $CX_{1\to2}$ copies all dependences of $q_1$ to $q_2$, and another $CX_{1\to2}$ will attempt to copy all the same dependences again from $q_1$ to $q_2$. As $CX_{1\to2}$ is its own inverse, the 2$^{nd}$ $CX_{1\to2}$ thus cancels all dependences copied by the 1$^{st}$ $CX_{1\to2}$. A more interesting case happens when the dependence originated from the same unitary source is received from different qubits. For example, consider the 3-qubit state:

$$\left(a_1|0\rangle_1+a_2|1\rangle_1\right)\otimes\left(b_1|0\rangle_2+b_2|1\rangle_2\right)\otimes\left(c_1|0\rangle_3+c_2|1\rangle_3\right) \quad (4)$$

which starts with $q_1$ depending on $(a_1, a_2)$, $q_2$ depending on $(b_1, b_2)$, $q_3$ depending on $(c_1, c_2)$. Now if we apply $CX_{2\to1}$ and $CX_{2\to3}$, it becomes:

$$\left(a_1|0\rangle_1 + a_2|1\rangle_1\right) \otimes b_1|0\rangle_2 \otimes \left(c_1|0\rangle_3 + c_2|1\rangle_3\right) + \left(a_2|0\rangle_1 + a_1|1\rangle_1\right) \otimes b_2|1\rangle_2 \otimes \left(c_2|0\rangle_3 + c_1|1\rangle_3\right) \quad (5)$$

where $q_1$ and $q_3$ have both acquired dependence on $(b_1, b_2)$. Now if we apply $CX_{1\to3}$, then the state becomes:

$$
\begin{aligned}
&|0\rangle_1 \otimes \left[ a_1 b_1 |0\rangle_2 \otimes \left(c_1|0\rangle_3 + c_2|1\rangle_3\right) + a_2 b_2 |1\rangle_2 \otimes \left(c_2|0\rangle_3 + c_1|1\rangle_3\right) \right] \\
&+ |1\rangle_1 \otimes \left[ a_2 b_1 |0\rangle_2 \otimes \left(c_2|0\rangle_3 + c_1|1\rangle_3\right) + a_1 b_2 |1\rangle_2 \otimes \left(c_1|0\rangle_3 + c_2|1\rangle_3\right) \right]
\end{aligned}
\quad (6)
$$

where $q_3$ has lost the dependence on $(b_1, b_2)$ because after some algebra we find that $p(|0\rangle_3) = |a_1 c_1|^2 + |a_2 c_2|^2$. Here the effect of $CX_{1\to3}$ is attempting to copy both $(a_1, a_2)$ and $(b_1, b_2)$ dependences from $q_1$ to $q_3$, however because $q_3$ has already received $(b_1, b_2)$ from $q_2$, receiving the same dependence again from $q_1$ will duplicate and cancel it on $q_3$. Here we propose the general cancelling rule of unitary dependences.

> **Definition 2**: Multiple dependences are considered "the same" if they are copied from the same qubit $q_i$ by some CNOT gates and there are no 1-qubit unitaries in between these CNOT gates. Furthermore, these dependences stay the same no matter how many times they are copied by more CNOT gates and which qubits they are on, as long as there are no 1-qubit unitaries applied.

> **Rule 3**: If a qubit $q_k$ receives the same dependence twice, it will lose (cancel) that dependence.

**Proof for Rule 3**: As discussed in the proof for Rule 2, $CX_{i\to j}$ calculates the binary sum $q_i \oplus q_j$ and stores its value on $q_j$, and if we next apply e.g. $CX_{j\to k}$ then it calculates the binary sum $q_i \oplus q_j \oplus q_k$ and stores its value on $q_k$. Note as long as there are no 1-qubit unitaries and only CNOTs, we can continue to perform such binary additions with more CNOT gates, so as to update the current "configuration" represented by an account of the current value held by each qubit. This process can be clearly described by the "qubit functional configuration" (QFC) where we start with the initial QFC (for details of the QFC theory see Ref. [25]):

$$(f_1 = q_1, f_2 = q_2, f_3 = q_3..., f_n = q_n) \quad (7)$$

for which each $f_k$ (i.e. the qubit functional on $q_k$) represents the current value stored on the corresponding qubit $q_k$. Now if we apply e.g. $CX_{1\to2}$, we update $f_2$ to be $f_2 = q_1 \oplus q_2$ with other $f_k$s intact so the QFC becomes:

$$\left(f_1 = q_1, f_2 = q_1 \oplus q_2, f_3 = q_3..., f_n = q_n\right) \tag{8}$$

Now if we apply another CNOT gate e.g. $CX_{2\to3}$, we update $f_3$ to be $f_3 = q_1 \oplus q_2 \oplus q_3$ and the QFC becomes:

$$\left(f_1 = q_1, f_2 = q_1 \oplus q_2, f_3 = q_1 \oplus q_2 \oplus q_3..., f_n = q_n\right) \tag{9}$$

where we see that although $q_1$ has not directly interacted with $q_3$, it has nonetheless been connected to $q_3$ by the two CNOT gates and the connection is clear from the QFC where $f_1$ and $f_3$ share the same component of $q_1$. Here we notice the interesting fact that, the connection of $f_1$ and $f_3$ sharing the same component of $q_1$ also means $q_1$ and $q_3$ sharing the dependences that already exist on $q_1$ before the $CX_{1\to2}$ gate. This is clear because by Rule 2 above, $CX_{1\to2}$ copies all dependence from $q_1$ to $q_2$, and then $CX_{2\to3}$ copies all dependence from $q_2$ to $q_3$ – we see that the process of the qubit functionals getting updated by some CNOT gates is essentially the same as the process of the unitary dependences getting copied by the same CNOT gates. In other words, take Equation (9) as an example, $f_2 = q_1 \oplus q_2$ means $q_2$ shares the dependences initially on $q_1$, while $f_3 = q_1 \oplus q_2 \oplus q_3$ means $q_3$ shares the dependences initially on $q_1$ and $q_2$. With the equivalence between the two processes understood, it becomes clear why duplicated dependences get cancelled: no matter from which qubit $q_k$ receives the 1st copy of $q_i$'s initial dependences, its corresponding functional $f_k$ now has $q_i$ as a component, and then the attempt to add a 2nd copy of $q_i$'s initial dependences will add $q_i$ to $f_k$ again, and this cancels $q_i$ on $f_k$ because $q_i \oplus q_i = 0$. In addition, it also becomes clear why the cancelling requires the condition of no 1-qubit unitaries, because only then can we stay in the same QFC addition process or as described in Ref. [25] "in the same QFC layer", while any 1-qubit unitary will require the QFC getting reset to the initial configuration. In terms of the unitary dependence theory, if $q_k$ receives the 1st copy of $q_i$'s dependences from $q_j$ after a 1-qubit unitary $U_j$ on $q_j$, the dependences could be modified by $U_j$ such that they are no longer the same as the initial ones. Now when $q_k$ receives the 2nd copy from $q_h$ before any 1-qubit unitary $U_h$ on $q_h$, then this copy stays the same as the initial one, thus the two copies of dependences may not cancel. This concludes the proof for Rule 3.

The QFC description of quantum circuits and states not only helps us prove Rule 3, but can also be useful for identifying connections between qubits with no apparent interactions. In Definition 2 we have the fact that "dependences stay the same no matter how many times they are copied by more CNOT gates and which qubits they are on, as long as there are no 1-qubit unitaries applied". This means two qubits may acquire the same dependences without ever interacting directly and they will be dependent when manipulated or measured. In addition, cancelling of those dependences will happen when some time later the two qubits directly interact with a CNOT gate.

**2.3.2** To illustrate the application of all three rules working together to obtain the complete unitary dependence picture, here we consider the hardware-efficient ansatz that is widely used in variational quantum algorithms [30-33]. The main feature of the hardware-efficient ansatz is that each ansatz layer includes two sub-layers: one sub-layer of parameterized 1-qubit unitaries and one sub-layer of two-qubit entanglers. When the two-qubit entanglers are all CNOT gates, there is no 1-qubit unitary in between the entanglers and the entangler sub-layer can be considered as a single layer of the qubit functional configuration (QFC) as described above – this means the cancelling rule works perfectly within one entangler sub-layer and we can easily obtain the dependence picture of each ansatz layer. In two examples, the ansatz used in Ref. [32] and Ref. [33] are shown in Figure 5 and Figure 6 respectively, and the unitary dependence pictures (in Equation (3)'s description form of the picture) are shown in Equations (10) and (11) respectively.

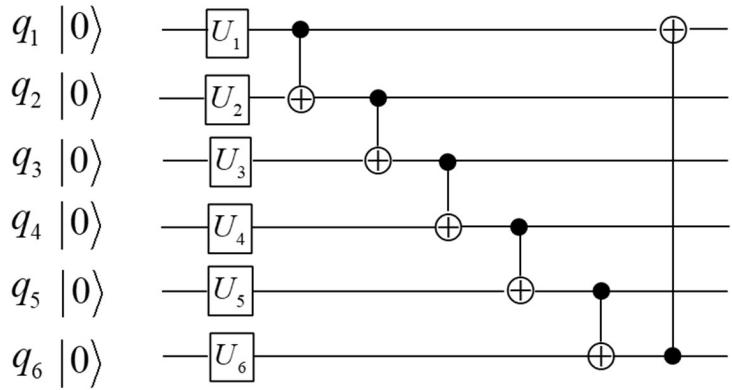

Figure 5. Showing the creation circuit of one layer of the hardware-efficient ansatz in Ref. [32].

The unitary dependence picture of the ansatz in Ref. [31]

$$(q_1:\{U_2 \sim U_6\}, q_2:\{U_1,U_2\}, q_3:\{U_1 \sim U_3\}, q_4:\{U_1 \sim U_4\}, q_5:\{U_1 \sim U_5\}, q_6:\{U_1 \sim U_6\}) \quad (10)$$
$$(U_1:\{q_2 \sim q_6\}, U_2:\{q_1 \sim q_6\}, U_3:\{q_1,q_3 \sim q_6\}, U_4:\{q_1,q_4 \sim q_6\}, U_5:\{q_1,q_5,q_6\}, U_6:\{q_1,q_6\})$$

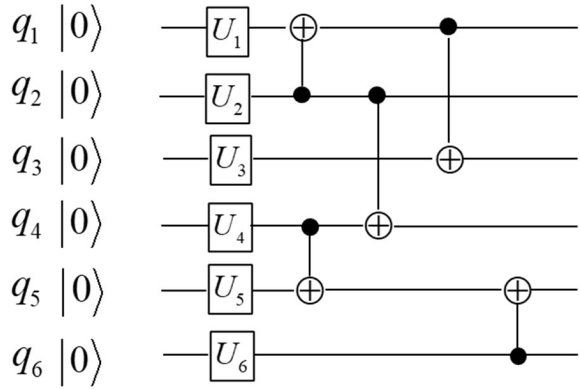

Figure 6. Showing the creation circuit of one layer of the hardware-efficient ansatz in Ref. [33].

The unitary dependence picture of the ansatz in Ref. [32]

$$\left(q_1:\{U_1,U_2\},\ q_2:\{U_2\},\ q_3:\{U_1 \sim U_3\},\ q_4:\{U_2,U_4\},\ q_5:\{U_4 \sim U_6\},\ q_6:\{U_6\}\right) \quad (11)$$

$$\left(U_1:\{q_1,q_3\},\ U_2:\{q_1 \sim q_4\},\ U_3:\{q_3\},\ U_4:\{q_4,q_5\},\ U_5:\{q_5\}, U_6:\{q_5,q_6\}\right)$$

In Equations (10) and (11) we see the differences between the ansatzes in Figure 5 and Figure 6 are clearly shown by the unitary dependence pictures. For example, in Equation (11) there are qubits ($q_2$ and $q_6$) that only have one unitary dependence and unitaries that only belong to one qubit ($U_3$ and $U_5$), while in Equation (10) there are no such qubits or unitaries. Therefore we immediately know that the measurement and manipulation of the qubits are more dependent in Figure 5 and more independent in Figure 6 – thus the two ansatzes behave very differently. This demonstrates the ability of the unitary dependence theory to characterize the behaviors of different ansatzes.

**2.3.3** Note if two copies of dependences are not considered the same by Definition 2, they may or may not cancel when copied to the same qubit. Consider the simple example of a 2-qubit state:

$$|\phi\rangle = \left(a_1|0\rangle_1 + a_2|1\rangle_1\right) \otimes \left(b_1|0\rangle_2 + b_2|1\rangle_2\right) \quad (12)$$

with $q_1$ depending on $(a_1,a_2)$ and $q_2$ depending on $(b_1,b_2)$. Now applying $CX_{1\to2}$ we have:

$$|CX_{1\to2}\phi\rangle = a_1|0\rangle_1 \otimes \left(b_1|0\rangle_2 + b_2|1\rangle_2\right) + a_2|1\rangle_1 \otimes \left(b_2|0\rangle_2 + b_1|1\rangle_2\right) \quad (13)$$

and by Rule 2, $q_2$ now depends on both $(b_1,b_2)$ and $(a_1,a_2)$ with $p(|0\rangle_2) = |a_1 b_1|^2 + |a_2 b_2|^2$. By Rule 3, if we apply $CX_{1\to2}$ again then $(a_1,a_2)$ will be canceled from $q_2$, but if we apply a 1-qubit unitary $U_2 = \begin{pmatrix} u_1 & u_2^* \\ u_2 & -u_1^* \end{pmatrix}$ on $q_2$ first, then $CX_{1\to2}$, we have:

$$|CX_{1\to2} \cdot U_2 \cdot CX_{1\to2}\phi\rangle = a_1|0\rangle_1 \otimes \left[\left(u_1 b_1 + u_2^* b_2\right)|0\rangle_2 + \left(u_2 b_1 - u_1^* b_2\right)|1\rangle_2\right]$$
$$+ a_2|1\rangle_1 \otimes \left[\left(u_2 b_2 - u_1^* b_1\right)|0\rangle_2 + \left(u_1 b_2 + u_2^* b_1\right)|1\rangle_2\right] \quad (14)$$

and the probability of measuring $|0\rangle$ for $q_2$ is:

$$p(|0\rangle_2) = |a_1|^2 \left|u_1 b_1 + u_2^* b_2\right|^2 + |a_2|^2 \left|u_2 b_2 - u_1^* b_1\right|^2 \quad (15)$$

which still has dependence on $(a_1,a_2)$, i.e. cancelling did not happen. This is clearly due to $U_2$'s modification of the 1st copy of $(a_1,a_2)$ received by $q_2$.

However, a more intriguing scenario happens if we entangle a 3$^{rd}$ qubit $q_3 = |0\rangle_3$ to $q_2$ by applying $CX_{2\to3}$ to the state $|CX_{1\to2}\phi\rangle$ as in Equation (13), we have:

$$|CX_{2\to3} \cdot CX_{1\to2}\phi\rangle = a_1|0\rangle_1 \otimes (b_1|00\rangle_{23} + b_2|11\rangle_{23}) + a_2|1\rangle_1 \otimes (b_2|00\rangle_{23} + b_1|11\rangle_{23}) \quad (16)$$

where by Rule 2 we have copied $q_2$'s dependences to $q_3$, while there is no apparent change to $q_1$ and $q_2$. If we now apply $U_2 = \begin{pmatrix} u_1 & u_2^* \\ u_2 & -u_1^* \end{pmatrix}$ on $q_2$ and then $CX_{1\to2}$, we have:

$$\begin{aligned}
&|CX_{1\to2} \cdot U_2 \cdot CX_{2\to3} \cdot CX_{1\to2}\phi\rangle \\
&= a_1|0\rangle_1 \otimes \left[ b_1(u_1|0\rangle_2 + u_2|1\rangle_2) \otimes |0\rangle_3 + b_2(u_2^*|0\rangle_2 - u_1^*|1\rangle_2) \otimes |1\rangle_3 \right] \\
&+ a_2|1\rangle_1 \otimes \left[ b_2(u_2|0\rangle_2 + u_1|1\rangle_2) \otimes |0\rangle_3 + b_1(-u_1^*|0\rangle_2 + u_2^*|1\rangle_2) \otimes |1\rangle_3 \right]
\end{aligned} \quad (17)$$

and the probability of measuring $|0\rangle$ for $q_2$ is:

$$\begin{aligned}
p(|0\rangle_2) &= |a_1|^2 (|u_1 b_1|^2 + |u_2^* b_2|^2) + |a_2|^2 (|u_2 b_2|^2 + |u_1^* b_1|^2) \\
&= |u_1 b_1|^2 + |u_2 b_2|^2
\end{aligned} \quad (18)$$

which has no dependence on $(a_1, a_2)$. So although Rule 3's condition is violated by $U_2$ before $CX_{1\to2}$, cancelling still happens! By comparing Equations (13) with (16), (14) with (17), we see the only difference between the two is, in the 2$^{nd}$ case, $q_2$ is entangled to $q_3$ before $U_2$, and this suggests that *entanglement can protect the cancellability from getting broken by local 1-qubit unitaries*. To understand this interesting phenomenon, we observe the state in Equation (13) $q_1$ and $q_2$ are entangled, such that a local unitary $U_2$ on $q_2$ may affect the dependence on $(a_1, a_2)$ that is shared by the two qubits. On the other hand, if we trace out $q_3$ in Equation (16) to get the reduced density matrix of $q_1$ and $q_2$, we have:

$$\begin{aligned}
Tr_3(\rho) &= Tr_3(|CX_{2\to3} \cdot CX_{1\to2}\phi\rangle\langle CX_{2\to3} \cdot CX_{1\to2}\phi|) \\
&= (|a_1 b_1|^2 + |a_2 b_2|^2)|\varphi_1\rangle\langle\varphi_1| + (|a_1 b_2|^2 + |a_2 b_1|^2)|\varphi_2\rangle\langle\varphi_2| \\
|\varphi_1\rangle &= \frac{(a_1 b_1|0\rangle_1 + a_2 b_2|1\rangle_1)}{\sqrt{|a_1 b_1|^2 + |a_2 b_2|^2}} \otimes |0\rangle_2 \quad |\varphi_2\rangle = \frac{(a_1 b_2|0\rangle_1 + a_2 b_1|1\rangle_1)}{\sqrt{|a_1 b_2|^2 + |a_2 b_1|^2}} \otimes |1\rangle_2
\end{aligned} \quad (19)$$

where the reduced density matrix of $q_1$ and $q_2$ is a mixture of two pure product states $|\varphi_1\rangle$ and $|\varphi_2\rangle$, and the shared dependence on $(a_1, a_2)$ only affects $q_2$ through the pure state probabilities

$p_1 = |a_1 b_1|^2 + |a_2 b_2|^2$ and $p_2 = |a_1 b_2|^2 + |a_2 b_1|^2$. Clearly, any 1-qubit unitaries on $q_2$ will only affect the qubit locally and will not affect the pure state probabilities, and this is the reason why cancelling holds even after $U_2$. The effect of entanglement is to break the entangled state in Equation (13) into product states in the 2-qubit reduced system in Equation (19), and this causes the cancelling of dependences to happen even when local 1-qubit unitaries are applied to $q_2$. This example may suggest a mechanism for using entanglement with additional qubits to protect certain properties of the system against local disturbances.

### 3. Conclusion

In this work we develop a unitary dependence theory to characterize the behaviors of quantum circuits and states in terms of how 1-qubit unitaries and CNOT gates affect qubits and determine their measurement probabilities. In particular, we define the basic rules of dependence creation by 1-qubit unitaries and copy by CNOT gates: 1. a 1-qubit unitary $U_k$ makes the qubit $q_k$'s measurement probabilities depend on $U_k$; 2. a CNOT gate $CX_{j \to k}$ copies all the control qubit $q_j$'s dependences to the target qubit $q_k$. By these rules, after a gate sequence of a quantum circuit, the final state can be described by a complete dependence picture that shows which qubits depend on which 1-qubit unitaries. The dependence picture carries important information of whether the measurement results of qubits are dependent or independent, and whether multiple qubits can be manipulated together or separately. Compared to the abstract formalism of multi-qubit entanglement, the dependence picture is more directly connected to the practicalities of using parameterized quantum gates to manipulate qubits and create desirable measurement statistics in the output states. In addition, the dependence picture is easier to use for many-qubit systems and more robust upon system partitioning. Under certain conditions, the dependences originated from the same unitary source can cancel when duplicated on the same qubit, which reduces complexity and simplifies the dependence picture. A particularly interesting case arises when studying the cancellability of dependences is that entanglement with an additional qubit may protect the cancellability from getting broken by local 1-qubit unitaries. This may suggest a mechanism for using entanglement with additional qubits to protect certain properties of the system against local disturbances. Finally, the theory has been applied to the widely-used hardware-efficient ansatz to demonstrate its ability to characterize the behaviors of different ansatzes.

**Acknowledgement**

ZH and SK acknowledge funding by the U.S. Department of Energy (Office of Basic Energy Sciences) under Award No. DE-SC0019215.